\documentclass[10pt,conference]{IEEEtran}
\IEEEoverridecommandlockouts
\usepackage{cite}
\usepackage{amsmath,amssymb,amsfonts}
\usepackage{algorithm}
\usepackage[noend]{algorithmic}
\usepackage{graphicx}
\usepackage{textcomp}
\usepackage{xcolor}
\usepackage{bm}
\usepackage{url}
\usepackage{wrapfig}
\usepackage{eso-pic}
\def\BibTeX{{\rm B\kern-.05em{\sc i\kern-.025em b}\kern-.08em
    T\kern-.1667em\lower.7ex\hbox{E}\kern-.125emX}}

\newcommand\AtPageUpperMyright[1]{\AtPageUpperLeft{
 \put(\LenToUnit{0.02\paperwidth},\LenToUnit{-1cm}){
     \parbox{1.07\textwidth}{\raggedleft\fontsize{7}{11}\selectfont #1}}
 }}
\newcommand{\conf}[1]{
\AddToShipoutPictureBG*{
\AtPageUpperMyright{#1}
}
}

\begin{document}

\title{
    Dividing Deep Learning Model for Continuous Anomaly Detection of Inconsistent ICT Systems
}

\author{
\IEEEauthorblockN{Kengo Tajiri$^{\ast}$, Yasuhiro Ikeda$^{\dagger}$, Yuusuke Nakano$^{\ast}$, Keishiro Watanabe$^{\ast}$}
\IEEEauthorblockA{$^{\ast}$NTT Network Technology Laboratories, NTT Corporation, Tokyo 180-8585, Japan}
\IEEEauthorblockA{$^{\dagger}$PKSHA Technology, Tokyo 113-0033, Japan}
Email: $^{\ast}$\{kengo.tajiri.bk, yuusuke.nakano.dn, keishiro.watanabe.ry\}@hco.ntt.co.jp,
 $^{\dagger}$yasuhiro\_ikeda@pkshatech.com
 \vspace{-5mm}
}
\conf{ \copyright 2020 IEEE.  Personal use of this material is permitted.  Permission from IEEE must be obtained for all other uses, in any current or future media, including reprinting/republishing this material for advertising or promotional purposes, creating new collective works, for resale or redistribution to servers or lists, or reuse of any copyrighted component of this work in other works.} 

\maketitle

\begin{abstract}
    Health monitoring is important for maintaining reliable information and communications technology (ICT) systems.
    Anomaly detection methods based on machine learning, which train a model for describing "normality" are promising for monitoring the state of ICT systems.
    However, these methods cannot be used when the type of monitored log data changes from that of training data due to the replacement of certain equipment.
    Therefore, such methods may dismiss an anomaly that appears when log data changes.
    To solve this problem, we propose an ICT-systems-monitoring method with deep learning models divided based on the correlation of log data.
    We also propose an algorithm for extracting the correlations of log data from a deep learning model and separating log data based on the correlation.
    When some of the log data changes, our method can continue health monitoring with the divided models which are not affected by changes in the log data. 
    We present the results from experiments involving benchmark data and real log data, which indicate that our method using divided models does not decrease anomaly detection accuracy and a model for anomaly detection can be divided to continue monitoring a network state even if some the log data change. 
\end{abstract}


\section{Introduction}
Health monitoring is important for maintaining reliable information and communications technology (ICT) systems.
ICT systems have become more complex and contain more types of services.
Therefore, many types of data to be monitored are now output from ICT systems.
Since it is difficult to configure the monitoring policy for such complex data with the domain knowledge of operators, anomaly detection methods based on machine learning are promising for monitoring the state of ICT systems.
Anomaly detection methods, which train a model that describes "normality" from the correlation between the variables of normal data, are often used for monitoring ICT systems\cite{lakhina2004diagnosing,aygun2017network,ahmed2016survey} because of the imbalance between the amount of normal data and that of anomaly data and the difficulty of covering all types of anomaly data \cite{chalapathy2019deep}.

Learning normal correlations of data in such methods is often achieved through dimensionality reduction for the feature vectors transformed from raw data.
Correlating features can be represented by fewer latent features and reducing the dimensions of the feature vectors, therefore, corresponds to learning the correlations among the dimensions.
The feature vector of test data is evaluated according to if the dimensionality reduction by the model trained with only normal data succeeds.
Since calculating dimensionality reduction involves all dimensions of feature vectors, the dimensions of feature vectors should be consistent during both training and test.

Let us consider anomaly detection for ICT systems.
Since the pieces of equipment making up an ICT system are connected, the normal correlation of log data from not a piece of equipment but the entire system should be examined.  
However, hardware may be added, reduced, or replaced, and the configuration of software may be changed during system maintenance in ICT systems and the dimensions of the transformed feature vectors may change since some of the log data from the system change in such a case.
If the dimensions change, the trained models for the systems cannot be used for the new data from the systems.
Therefore, ICT systems cannot be monitored when collecting new data and training a new model.
If a failure occurs during that time, operators may not notice it, which may lead to serious problems. 

To ease the constraint of anomaly detections, we propose an ICT-system-monitoring method for continuous anomaly detection even if the dimensions of the vectors change. 
In this method, the dimensions of the feature vectors are separated based on the correlation between them.
Then, models for anomaly detection are built for each group of dimensions.
Dividing a model for anomaly detection based on the correlation prevents the changes in log data occurring due to local changes in ICT systems from affecting the entire model.
Taking into account the non-linear nature of the monitored data, we leverage deep neural networks for both separating the dimensions and building anomaly detection models.
We also propose an algorithm for extracting correlations among the dimensions from the black-box neural network and separating the dimensions based on the correlation.

We now discuss related work on calculating correlation and anomaly detections.
There are two types of methods for calculating the correlation of dimensions: linear and non-linear.
Linear correlation is calculated as a Pearson correlation coefficient \cite{swinscow2002statistics}.
Non-linear correlation is calculated as a Spearman correlation coefficient \cite{swinscow2002statistics} or, maximal information coefficient \cite{reshef2011detecting} or calculated by the Hilbert-Schmidt independence criterion \cite{gretton2005measuring}.
Since these methods are used to calculate the correlation between two dimensions, they require $M(M-1)/2$ calculations for the number of dimensions of a feature vector: $M$.
The number of calculations increases explosively for large dimensional data.
As discussed in the following section, our method can be used to examine the correlation among all dimensions at once by training deep learning models without examining all combinations with training data in order.

Since principal component analysis (PCA) \cite{jolliffe2011principal} and autoencoder (AE) \cite{hinton2006reducing} can learn the correlation between the dimensions of feature vectors through dimensionality reduction, both are used as anomaly detection methods, which train a model that describes normality. 
PCA reduces the dimensions based on linear correlation, whereas AE, which is a deep learning based method, can reduce the dimensions based on non-linear correlation \cite{sakurada2014anomaly}.
Thus, an AE can detect anomaly data based on a more complex relation between the dimensions than PCA.


\section{Proposed method}
Our method consists of two parts: separating dimensions and building models for anomaly detection.
In both parts, an AE is used because considering non-linear correlation is important for the complex log data from ICT systems, which have continuous values, such as the use rate of CPUs and network bandwidth, and discrete values such as the number of syslog occurrences.
We first explain an AE and an anomaly detection method based on the AE.
Then, we explain the proposed algorithm for separating dimensions of feature vectors.

\subsection{Autoencoder}\label{AE}
An AE model has two parts of neural networks: encoder and decoder.
The encoder $\bm{E}(\cdot)$ compresses each input vector $\bm{x}$ into a low dimensional representation $E(x)$, and the decoder $\bm{D}(\cdot)$ reconstructs each $\bm{x}$ from $\bm{E}(\bm{x})$ as $\bm{D}(\bm{E}(\bm{x}))$.
An AE model is optimized to minimize the following loss function.
\begin{equation}
    \label{AE_loss}
    \mathcal{L}_{\rm AE}(\bm{x}) = \frac{1}{M}||\bm{x}-\bm{D}(\bm{E}(\bm{x}))||_2 +\lambda \Sigma_{l=1}^{N-1}||W^{(l)}||_p,
\end{equation}
where $||\cdot||_p$ is the $L_p$-norm.
The second term is a regularization term to suppress over-fitting for training data and $\lambda$ is a hyperparameter that determines the magnitude of regularization.

\subsection{Anomaly detection with AE}
Since an AE model is trained to reconstruct training data, input similar to training data can be reconstructed well with the trained AE model.
Whereas, the reconstruction error of input, which differs from that of the training data becomes large.
Thus, the reconstruction error of anomaly data is larger than that of normal data.
The reconstruction error, which corresponds to the first term of (\ref{AE_loss}), can be assumed as the anomaly score of input.
After the anomaly score of each test data point is calculated, each data point is classified as normal or anomaly data with the threshold for the anomaly score. 

\subsection{Extracting correlations and separating dimensions}
An AE model learns the correlation between the dimensions of training data to obtain $\bm{E}(\bm{x})$.
However, it is difficult to know which dimensions are correlated from the AE model, which is a black-box deep learning model.
Although many interpretation models have been proposed \cite{guidotti2018survey}, there has been no algorithm for extracting the correlations among the dimensions from an AE model.
Therefore, based on our hypothesis that when a dimension affects the reconstruction of another dimension in the AE model these dimensions correlate with each other, we leveraged DeepLIFT\cite{shrikumar2017learning}, an algorithm for quantifying the influence of input for output, for extracting these correlations.

DeepLIFT is an algorithm for calculating the importance score: $C_{\Delta x_i \Delta y}$ of each dimension $i$ of input $\bm{x}$ for output result $\bm{y}$ with the difference between original input $\bm{x}$ and reference input $\bm{x}_\mathrm{ref}$.  
The $\bm{x}_\mathrm{ref}$ can be determined arbitrarily, and we define this as zero vectors with the proposed algorithm. 
This interpretation algorithm can calculate more interpretable importance scores than other gradient based algorithms\cite{bach2015pixel,selvaraju2017grad}, because it takes into account the global shift in dimensions of input.
DeepLIFT was used for interpreting the classification results for the test data of MNIST and Genomics in \cite{shrikumar2017learning}.
Since we want to know the correlation between dimensions of training data with our proposed algorithm, we calculate the importance score of each dimension of input $x_i$ for each dimension of the output of the trained AE model: $x'_j = D(\bm{E}(\bm{x}))_j$.
The importance score becomes a $M\times M$ dimensional matrix as follows.
\begin{align}
    \label{importance_score_matrix}
    &Importance\ score = \nonumber\\
        &\left(
            \begin{array}{cccc}
              C_{\Delta x_1 \Delta x'_1} & C_{\Delta x_1 \Delta x'_2} & \ldots & C_{\Delta x_1 \Delta x'_M} \\
              C_{\Delta x_2 \Delta x'_1} & C_{\Delta x_2 \Delta x'_2} & \ldots & C_{\Delta x_2 \Delta x'_M} \\
              \vdots & \vdots & \ddots & \vdots \\
              C_{\Delta x_M \Delta x'_1} & C_{\Delta x_M \Delta x'_2} & \ldots & C_{\Delta x_M \Delta x'_M}
            \end{array}
        \right)
\end{align}

Each column of the matrix represents the magnitude of the effect of each dimension of input for a dimension of output.
The importance score is then averaged over the training data by each matrix element.
If the absolute value of an element of the matrix is larger than a predetermined threshold, the element is replaced with 1; otherwise, with 0.
The matrix is divided into column vectors, and the sets of column vectors are grouped by column vectors that are not orthogonal to one another.
The column vectors in the same group indicate that these dimensions of output are reconstructed with the same input dimensions.
Therefore, the dimensions of output in the same groups are assumed to be correlated with each other.

Finally, the dimensions of the training data are separated based on the group.
For each group, an AE model for anomaly detection is built and trained again.
For each test data point, the anomaly score is calculated with AE models.
If some AE models cannot be used, the reconstruction error is calculated only for the dimensions input to the available AE models.

We show the pseudo code of the algorithm for separating dimensions in Alg.\ref{alg1}.

\begin{algorithm}                      
    \caption{Proposed algorithm for separating dimensions \protect\linebreak
    (\# means the number of elements in a set.)}
    \label{alg1}                          
    \begin{algorithmic}[1]                  
    \REQUIRE $X_\mathrm{train}:$training data, $\tau:$threshold,\\
            $\bm{D}(\bm{E}(\cdot)):$trained AE model
    \ENSURE $G$: sets of groups
    \STATE $G$ := \{\}
    \STATE $R$ := \{\} (sets of column vectors)
    \STATE $S_{\Delta x, \Delta x'}$ := 0 ($\in \mathbb{R}^{M\times M}$)
    \STATE cnt := 1 
    \STATE calculate (\ref{importance_score_matrix}) with $X_\mathrm{train}$ and $\bm{D}(\bm{E}(\cdot))$.
    \STATE $C_{\Delta x_i, \Delta x'_j}$ is averaged over training dataset.
    \IF{$|C_{\Delta x_i, \Delta x'_j}| > \tau$}
    \STATE $S_{\Delta x_i, \Delta x'_j} \leftarrow 1$
    \ENDIF
    \STATE $S'_{\Delta x, \Delta x'}$ is considered as sets of vectors: $\bm{S}_{\Delta x, \Delta x_i'}$.
    \WHILE{\#$S'_{\Delta x, \Delta x'} > 0$}
    \STATE an $\bm{S}_{\Delta x, \Delta x'_i}$ is chosen and deleted from $S'_{\Delta x, \Delta x'}$.
    \IF{$\bm{S}_{\Delta x, \Delta x'_i} = \bm{0}$}
    \STATE $i$ is input to $g_0$.
    \IF{$g_0 \notin G$}
    \STATE $g_0$ is input to $G$.
    \ENDIF
    \ELSE
    \STATE $i$ is input to $g_\mathrm{cnt}$.
    \STATE $\bm{r}_\mathrm{cnt}\leftarrow \bm{S}_{\Delta x, \Delta x'_i}$
    \STATE $g_\mathrm{cnt}$ is input to $G$.
    \STATE $\bm{r}_\mathrm{cnt}$ is input to $R$.
    \FORALL{$\bm{S}_{\Delta x, \Delta x'_j} \in S'_{\Delta x, \Delta x'}$}
    \IF{$\bm{S}_{\Delta x, \Delta x'_j}\cdot \bm{r}_\mathrm{cnt} \neq 0$}
    \STATE $\bm{r}_\mathrm{cnt}\leftarrow \bm{r}_\mathrm{cnt} + \bm{S}_{\Delta x, \Delta x'_j}$
    \STATE $j$ is input to $g_\mathrm{cnt}$
    \STATE $\bm{S}_{\Delta x, \Delta x'_j}$ is deleted from $S'_{\Delta x, \Delta x'}$.
    \ENDIF
    \ENDFOR
    \STATE $\mathrm{cnt}\leftarrow \mathrm{cnt} + 1$
    \ENDIF
    \ENDWHILE
    \WHILE{\#$R>0$}
    \STATE an $\bm{r}_i$ is chosen and deleted from $R$.
    \FORALL{$\bm{r}_j \in R$}
    \IF{$\bm{r}_i\cdot \bm{r}_j \neq 0$}
    \STATE $g_i \leftarrow g_i \cup g_j$
    \STATE $g_j$ is deleted from $G$.
    \STATE $\bm{r}_j$ is deleted from $R$.
    \ENDIF
    \ENDFOR
    \ENDWHILE
    \end{algorithmic}
\end{algorithm}
\setlength{\textfloatsep}{6pt}
\section{Evaluation}
We evaluated the proposed method and algorithm with the benchmark data of network intrusion detection and real log data from our virtual network environment, respectively.
We compared the area under receiver operating characteristic (AUROC) calculated using our divided AE models with that calculated using one AE model to show our method does not decrease the AUROC of anomaly detection with benchmark data.
We also evaluated the proposed algorithm with real log data to confirm if it could separate sets of correlated dimensions qualitatively.
Although we did not evaluate anomaly detection accuracy with real log data since the data under anomalous situations could not be obtained, anomaly detection accuracy should be evaluated in future work.  
Table \ref{hypara} shows the hyperparameters used in the two experiments.

\begin{table}[tbp]
    \centering
    \caption{ Hyperparameters of AE models used in two experiments. In experiment with benchmark, we used same hyperparameters for separation and anomaly detection. \label{hypara}}
    \vspace{-2mm}
    \begin{tabular}{|l||c|c|} \hline
       & benchmark data & real log data \\ \hline \hline
      \# of epoch & 200 & 1000 \\ \hline
      \# of hidden layers & 1 & 1 \\ \hline
      ratio of dimensionality reduction & &\\
      (round up after decimal point of & 0.25 & 0.25\\ 
      \# of dims of hidden layers) & & \\ \hline
      activation function & LeakyRelu\cite{maas2013rectifier} & LeakyRelu \\ 
      (only use first layer) & ($\alpha=0.2$) & ($\alpha=0.2$) \\ \hline
      $\lambda$ (regularization term) & 5e-5 ($p$=1) & 5e-5 ($p$=1) \\ \hline
      batch size & 1024 & 16 \\ \hline
      dropout & None & None \\ \hline
      optimizer & Adam\cite{kingma2014adam} & Adam \\ \hline
      learning rate & 1e-3 & 1e-3 \\ \hline
    \end{tabular}
\end{table}

\subsection{Evaluation with network benchmark data}
We evaluated the proposed method with the NSL-KDD dataset \cite{nsl-kdd} to examine how dividing an anomaly detection model affects the AUROC of anomaly detection. 

\subsubsection{Data}
The NSL-KDD dataset contains data for training (KDDTrain+) and test (KDDTest+), both of which contain normal and anomaly data.
Anomaly data means harmful flows such as U2R and R2L.
We used only normal data in KDDTrain+ as the training data. 
Each data point in the dataset has 34 continuous attributes and 7 discrete attributes.
By representing the discrete attributes as one-hot vectors, we transformed the data into a set of 122 dimensions vectors.

\subsubsection{Evaluation methods}
We first trained an AE model with the training data.
We then calculated the importance score with these data.
We separated the dimensions with a threshold for the importance score.
The dimensions that were independent of all the dimensions including themselves were grouped.
For each group, we trained new models for anomaly detection with the same training data again.
Finally, we calculated the AUROC of the test data with the trained models.
We separated the dimensions with certain thresholds to evaluate the dependence of the AUROC on separating dimensions.

\begin{table}[tbp]
    \centering
    \caption{ Experimental results with NSL-KDD data. }
    \label{AUROC result}
    \vspace{-2mm}
    \begin{tabular}{|c||c|c|} \hline
       & \# of models &  \\ 
      threshold ratio & (\# of dims of each model) & AUROC \\ \hline \hline
      0  & 1 (122) & 0.934 \\ \hline
      0.1  & 8 (26,1,1,1,12,1,1,79) & 0.937 \\ \hline
      0.3  & 11 (21,1,1,1,1,5,6,1,1,1,83) & 0.935 \\ \hline 
      0.5  & 12 (17,1,1,1,1,5,5,1,2,1,1,86) & 0.929 \\
      \hline 
      \multicolumn{2}{|c|}{randomly divide into 8 models} & 0.924 \\
      \hline
    \end{tabular}
\end{table}

\subsubsection{Evaluation results}
The results are listed in Table \ref{AUROC result}.
We used $(threshold\ ratio)*(variance\ of\ importance scores)$ as the threshold.
The second column shows the numbers of models and dimensions of each model with each threshold.
As the threshold increases, the number of models increases because small importance score elements are ignored for a large threshold.
The second column shows that the number of dimensions in each model differs because the dependency between each dimension differs.
The last group has the most dimensions because the values in most dimensions are constant in the normal data in KDDTrain+.

The AUROC calculated with the proposed method are compared with that calculated with one AE model (threshold = 0).
Table \ref{AUROC result} shows that although the AUROC with AE models divided randomly is lower than that with one AE model, that with our method is not lower than that with one AE model (threshold = 0.1 or 0.3).
This indicates that the AE model for anomaly detection can be divided based on the correlation without decreasing anomaly detection accuracy.

\subsection{Evaluation with real log data}

\subsubsection{Data}
\begin{figure}[tbp]
    \centering
    \includegraphics[width=0.7\linewidth]{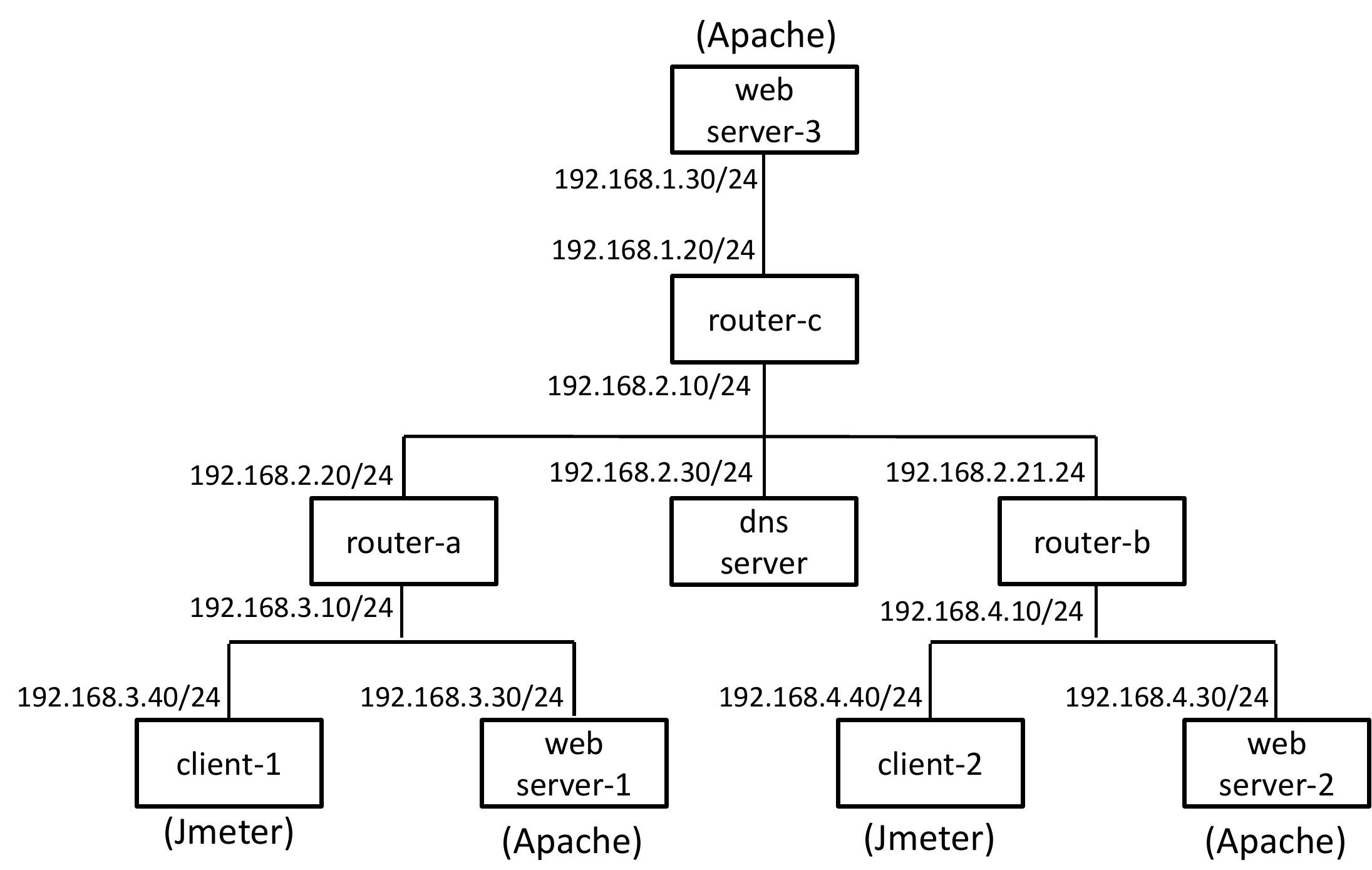}
    \vspace{-2mm}
    \caption{Topology of network used for collecting log data}
    \label{topology}
\end{figure}

We built a virtual network with virtual machines (VMs) and virtual LANs (VLANs) as shown in Fig.\ref{topology}.
Jmeter\cite{jmeter} was on each client VM and Apache HTTP Server was on each web server VM. 
Each client sent http requests to all of the web servers. 
The request frequencies from a client for all web servers were configured to be about the same values.
There were only http and DNS flows in this network.

We collected flows with L2 switches, MIB data from the web servers and routers with SNMP trap every minute, and syslogs from the web servers and routers during two hours.
To input these data for an AE model, we transformed them as the following.
First, the total value of packets (ipkt), that of bytes (ibyt), that of packets which did not take into account the difference of the client ports (count) were calculated for each source IP address (srcIP), destination IP address (disIP), protocol (ptc) and the port of the web servers or the dns server every minute.
The feature vectors of the flows had the following dimensions: "srcIP\_dstIP\_ptc\_(dns\textbar web server)port\_(count\textbar ipkt\textbar ibyt)".
Second, the information of diskIO, interfaceIO, and CPU of the web servers and routers were collected form the MIB data and transformed into feature vectors.
Third, the syslog appearances were counted every minute for each web server, router and syslog template ID, which were extracted using syslog templator\cite{kimura2018proactive}.
Finally, these feature vectors were joined.
The number of dimensions of the vectors was 259.
That from the flows was 36, that from the MIB data was 180 and that from the syslogs was 43.

\subsubsection{Evaluation methods}
We trained an AE model with all collected data, calculated the importance score and separated the dimensions with certain thresholds.
We qualitatively examined whether correlated dimensions can be grouped.

\subsubsection{Evaluation results}
The separation results were shown in Table \ref{separation result}.
When the threshold ratio was 2.0 and 6.0, the dimensions were separated into 8 and 13 groups, respectively.
The dimensions in the last group are independent of all of the dimensions in both separation results.
These results indicate that when the threshold ratio is large, the dimensions are finely separated, because the higher the threshold, the more severe the grouping decision.

When the threshold was 2, the 1st group was large and separating dimensions was not very successful.
However, when the threshold was 6, the flow data were separated with each client's IP address.
The reason is that since the frequencies of http request from client-1 and client-2 were different, the flow data from different srcIP were not correlated.
These results indicate that anomaly detection using flow data can continue when some of the log data change.
Contrary, the MIB data were separated based on the types of data rather than VMs. 
The reason is that since the frequencies of http request for all web servers from each client were about equal and the MIB data from routers and web servers were affected by the http requests from both clients, the MIB data from each piece of equipment were strongly correlated. 
However, since each model built with MIB data had only data from some of the VMs in this case, some models could continue anomaly detection even if some of the data changed.
Finally, we mention the last model, which had dimensions independent of all of the dimensions.
In this case, we grouped these independent dimensions.
However, since these dimensions are independent, they can be grouped arbitrarily.
If these dimensions are grouped for every piece of equipment, the continuity of anomaly detection can be enhanced.


\begin{table}[tbp]
    \centering
    \caption{ Experimental results with real log data. Each row indicates separated model. InterfaceIO, web server, router, and client are abbreviate as ifIO, srv, rt and clt, respectively. }
    \label{separation result}
    \vspace{-2mm}
    \begin{tabular}{|c||l|} \hline
      threshold & \\
      ratio & detail of model (\# of dim)\\ \hline \hline
      & http flow[all], CPU[srv1,2,3, rt-a,b,c], disk[srv2,3], \\
      & ifIO[srv1,2,3, rt-a,b,c], log("GET HTTP")[srv1,2,3] (88) \\ \cline{2-2}
      & dns flow[clt1$\rightarrow$ dnssrv] (3) \\ \cline{2-2}
      2& dns flow[clt2$\rightarrow$ dnssrv] (3) \\ \cline{2-2}
      & dns flow[dnssrv$\rightarrow$ clt1] (3) \\ \cline{2-2}
      & disk[srv1,3, rt-a,b,c], CPU[rt-c] (13) \\ \cline{2-2}
      & CPU[rt-c] (1) \\ \cline{2-2}
      & log("systemd") [srv1,2,3] (9)\\ \cline{2-2}
      & others (137) \\ \hline \hline
      & http flow[srv3$\rightarrow$ clt1,clt1$\rightarrow$ srv3,srv2$\rightarrow$ clt1,clt2$\rightarrow$ srv3] (12) \\ \cline{2-2}
      & http flow[srv3$\rightarrow$ clt2] (3) \\ \cline{2-2}
      & http flow[srv1$\rightarrow$ clt2] (3) \\ \cline{2-2}
      & http flow[clt1$\rightarrow$ srv2] (3) \\ \cline{2-2}
      & http flow[clt2$\rightarrow$ srv1] (3) \\ \cline{2-2}
      & dns flow[clt1$\rightarrow$ dnssrv] (3) \\ \cline{2-2}
      6& dns flow[clt2$\rightarrow$ dnssrv] (3) \\ \cline{2-2}
      & ifIO[srv2,3, rt-a,b,c] (26) \\ \cline{2-2}
      & CPU[srv2,3, rt-a,b,c] (4) \\ \cline{2-2}
      & CPU[rt-c] (1) \\ \cline{2-2}
      & diskIO[srv-1,3, rt-b,c] (10) \\ \cline{2-2}
      & log("GET HTTP")[srv1,2,3] (3) \\ \cline{2-2}
      & others (185) \\ \hline
    \end{tabular}
\end{table}

\section{Conclusion}
We proposed an ICT-system-monitoring method with AE models divided based on the correlation of monitored data and an algorithm for extracting the correlations among monitored data based on a deep learning model.
We showed that anomaly detection accuracy did not decrease when a model was divided based on the correlation with NSL-KDD data.
We also showed that real log data (i.e. flow, MIB and syslog) were grouped based on the characteristics of log data and our method could continue to monitor the network state with some AE models even if network equipment and log data changed.
For future work, we will evaluate anomaly detection accuracy with our method, when some of the data actually changes.

\bibliographystyle{unsrt}
\bibliography{neurips_2019}
\end{document}